\documentclass{article}
\usepackage{mrs2005,epsfig}
\usepackage[latin1]{inputenc}
\usepackage[T1]{fontenc}
\usepackage{multirow}
\usepackage{graphics}
\usepackage{afterpage}
\usepackage{graphicx}
\usepackage{subfigure}
\usepackage{lscape}
\usepackage{longtable}
\usepackage{supertabular}
\begin{document} 
\title{\textsc{Evolution of morphology in the Chandra Deep Field South}}

\author{Sébastien Lauger$^{1}$, Olivier Ilbert$^2$, Véronique Buat$^1$, Laurence Tresse$^1$, Denis Burgarella$^1$, Stéphane Arnouts$^1$, Olivier Le Fèvre$^1$, and the VVDS team$^{1, 2}$}
\affil{$^1$ Laboratoire d'Astrophysique de Marseille, Marseille, France \\$^2$ INAF-Osservatorio Astronomico di Bologna, Bologna, Italy}

\begin{abstract} 
We studied the morphology of galaxies in the Chandra Deep
Field South using ACS multi-wavelength data from the Great Observatories
Origin Deep Survey and 524 spectroscopic redshifts from the
VIMOS VLT Deep Survey  completed with 2874 photometric redshift computed 
from COMBO-17 multi-color data. The rest-frame B-band makes it
possible to discriminate two morphological types in an
asymmetry-concentration diagram: bulge- and disk-dominated
galaxies. 
  The rest-frame color index $B-I$ is found to be very correlated with the morphological classification: wholly bulge-dominated galaxies are redder than disk-dominated galaxies. However color allowed us to distinguish a population of faint blue bulge-dominated galaxies ($B-I<0.9$), whose nature is still unclear.  
  Using the rest-frame B-band classification from $z\sim0.15$ up to $z\sim1.1$, we quantified the evolution of the proportion of morphological types as a function of the redshift. Our large sample allowed us to compute luminosity functions per morphological type in rest-frame B-band. The bulge-dominated population is found to be composite: on the one hand the red ($(B-I)_{AB}>0.9$), bright galaxies, which seem to increase in density toward low redshifts. On the other hand the blue, compact, faint bulge-dominated galaxies, strongly evolving with the redshift.
\end{abstract} 
 
\section{Introduction} 
 Hierarchical models predict the formation of massive elliptical
galaxies by merging of smaller galaxies.
One way to put some constraints on these scenarios is to 
 measure the evolution of the morphology of galaxies with the
redshift 
to find a link with physical processes acting on
them.\\
The peculiar morphologies observed at high redshift (\textit{e.g.} \cite{van den bergh02}), 
and the need for automatic and objective morphological classification
led to the development of quantitative methods, notably based on
asymmetry and concentration parameters (\textit{e.g.}
\cite{bershady}, \cite{abraham96}, \cite{lauger}).
The morphological study of distant galaxies was most often restricted
to only one band, and their morphologies may suffer from morphological
$k$-corrections above $z\sim1$ and from the intrinsic galaxy evolution
(\cite{abraham96}, \cite{brinchmann98}). Today multi-band data provides a
rest-frame classification  comparable 
over all the redshift range. \\
Up to now, most of the luminosity functions (LFs) at low and high redshifts were computed using spectral or photometric types of galaxies. However a galaxy of a given morphological type may change of spectral type as a function of the redshift by following a passive evolution. Morphology brings a complementary, alternative way to study the evolution of galaxies. The automatic morphological classification of a large sample of galaxies allowed to compute the LFs per morphological type and per bin of redshift. The used cosmology is: $H_0=70$ km s$^{-1}$
Mpc$^{-1}$, $\Omega_0=0.3$, $\Lambda_0=0.7$. Magnitudes are given in the $AB$ system.

\section{Data and method overview}
\begin{figure}[htbp]
\vspace*{1.25cm}  
\centering
\begin{tabular}{cc}
\epsfig{figure=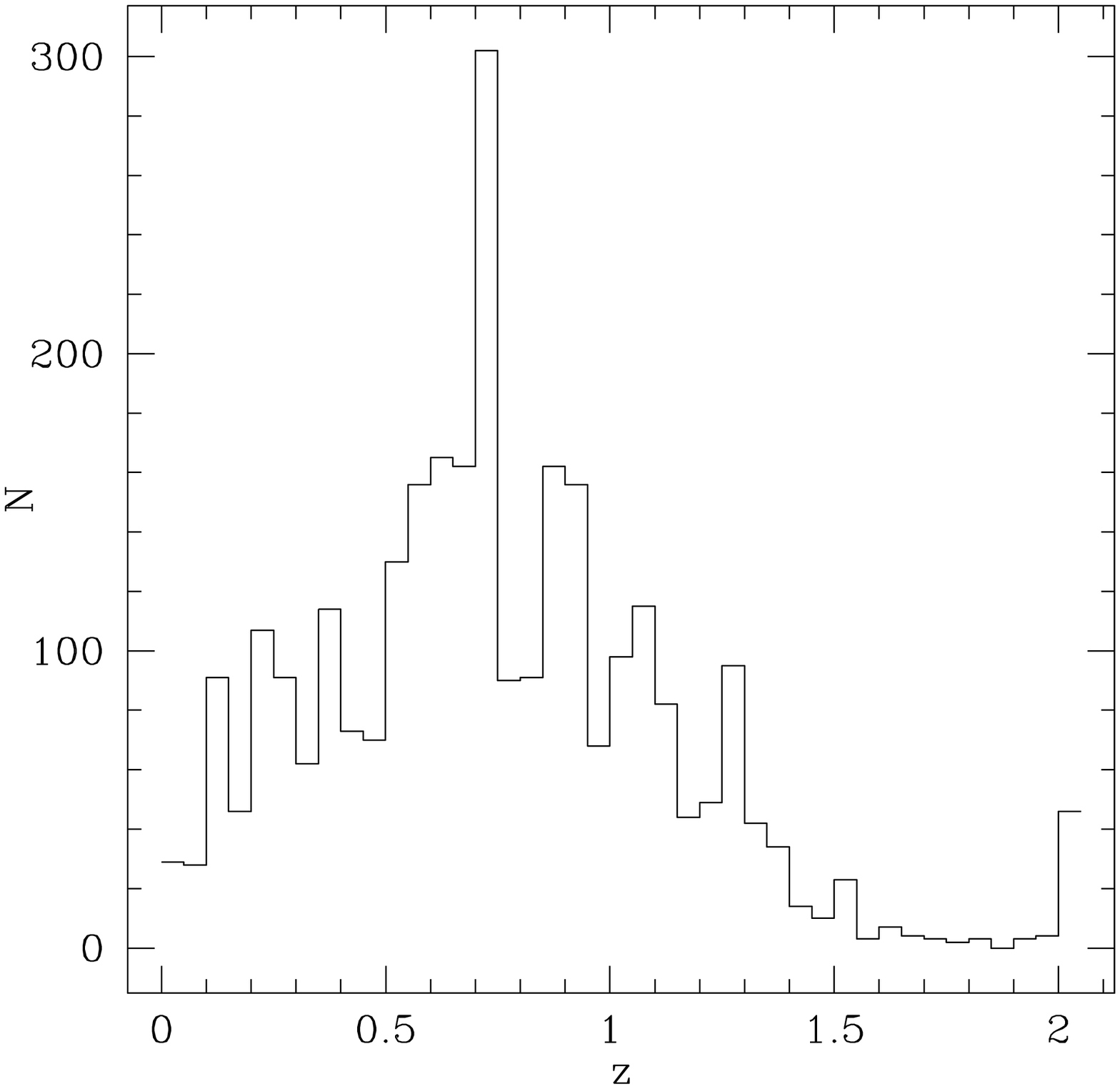,width=6.5cm} &  \epsfig{figure=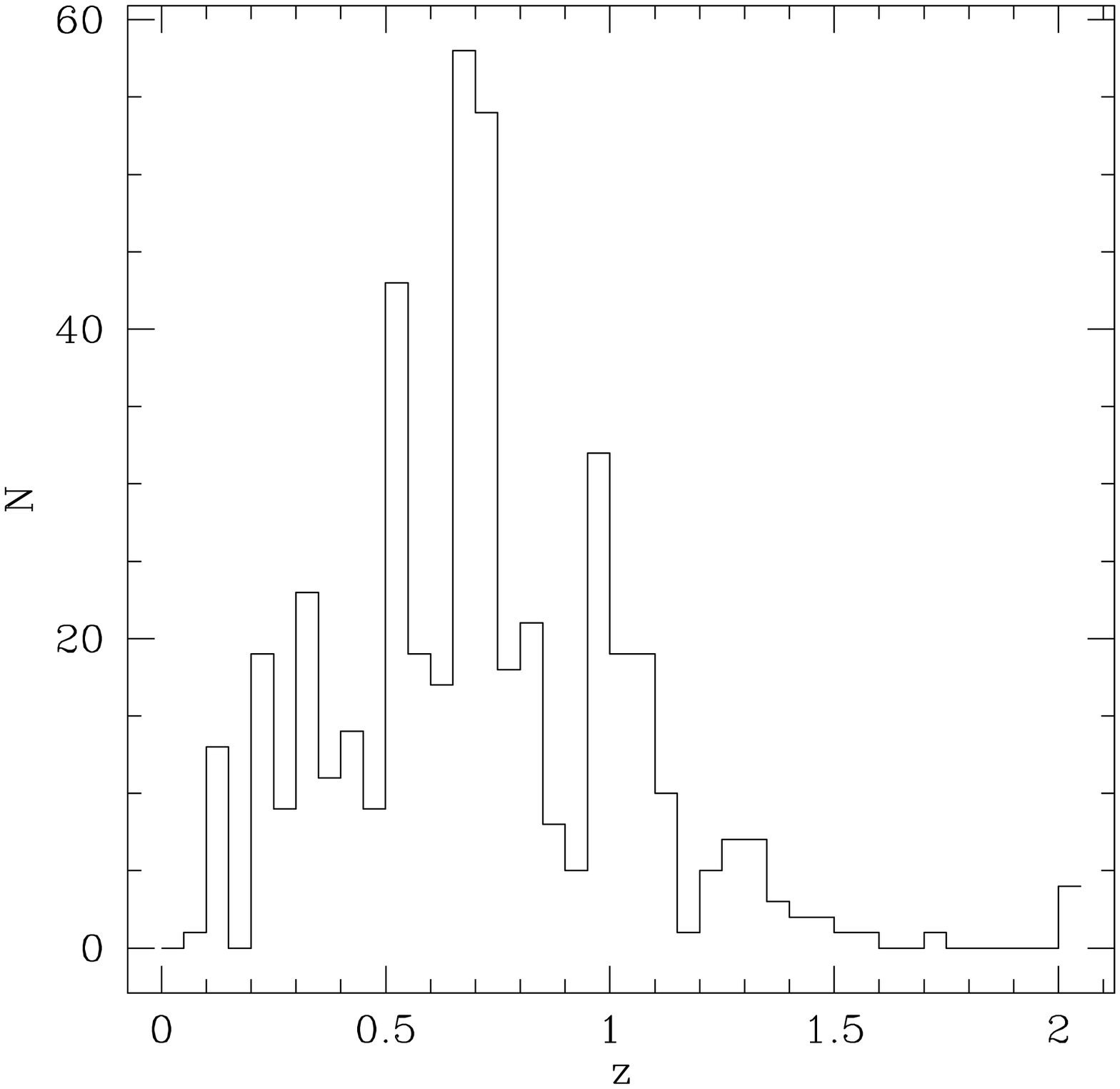,width=6.5cm}\\
\end{tabular}
\vspace*{0.25cm}  
\caption{Photometric (on the left) and spectroscopic (on the right) redshift distribution of the galaxies selected at $I_{AB}=24$.}
\label{fig: histo_z}
\end{figure}

We used images provided by the Great Observatories Origins Deep Survey
(GOODS, \cite{giavalisco04}) with the Hubble Space Telescope (HST) and
the Advanced Camera for Surveys (ACS) of the Chandra Deep Field South (CDFS)
 in the F850LP, F775W,
F606W and F435W filters (called hereafter $z$, $i$, $V$, and $B$
bands). From these images, we built two samples. \\ 
The first sample used the public data release of the VIMOS VLT Deep
Survey (VVDS, \cite{lefevre04}) which provide 1599 spectroscopic
redshifts up to $I_{AB}=24$ in the CDFS.  
So we secured HST multi-band imaging and redshifts for 524 galaxies up
to $I_{AB}=24$ and $z\sim4$. 
The second sample used the photometric  redshifts derived from
the multicolor catalogue of COMBO-17, described in \cite{ilbert}. This photometric redshift
  sample contains 2874 galaxies at $I_{AB}<24$. The redshift distribution of galaxies
shows that there is a dense structure around $z\sim0.7$ ({Figure
\ref{fig: histo_z})}. This is consistent with the results
 obtained with spectroscopic redshift (e.g. \cite{lefevre04}, \cite{Adami05}).\\
For both samples we computed concentration ($C$) and asymmetry ($A$) parameters
for the images of each galaxy in the four bands.  These parameters are
defined in \cite{lauger}.

\section{The morphological classification} 

\begin{figure}[htbp]
\vspace*{1.25cm}  
\centering
\epsfig{figure=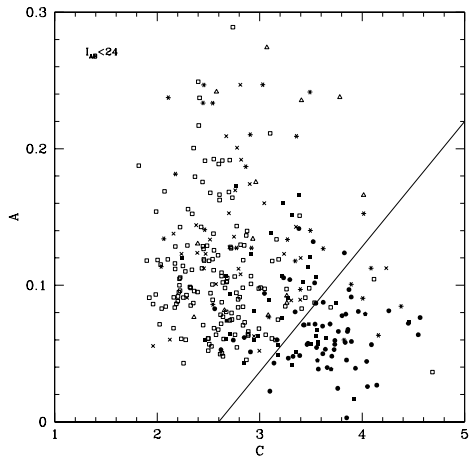,width=6.5cm}  
\vspace*{0.25cm}  
\caption{Asymmetry-concentration diagram of the spectroscopic redshift sample, as a function of the visual classification.}
\label{fig: AC eye}
\end{figure}

Although the method is subjective, we visually classified the galaxies of the spectroscopic redshift sample.
We defined the following visual classes: compact galaxies: quasi-stellar objects, small and concentrated.
 Elliptical galaxies: bulge-like galaxies without any outer visible structure.
 Early spiral galaxies: bulge-dominated galaxies with an outer visible disk.
 Late spiral galaxies: disk-dominated galaxies.
 Irregular galaxies: no apparent structure.
 Merger galaxies: major or minor disturbed appearance, double cores; and finally: miscallenous.\\
We plotted an $(A, C)$ diagram of the whole sample in rest-frame (RF) B-band as a
function of the visual classification (Figure \ref{fig: AC eye}). The visual classification is quite consistent with the
automatic classification. We visually traced a line separating the disk-dominated galaxies and the bulge-dominated
galaxies. Even if for the faintest galaxies the visual classification is strongly subjective and less reliable, the comparison between  quantitative and visual classifications remains in excellent  agreement.

\section{Morphology and color}

\begin{figure}[htbp]
\vspace*{1.25cm}  
\centering
\epsfig{figure=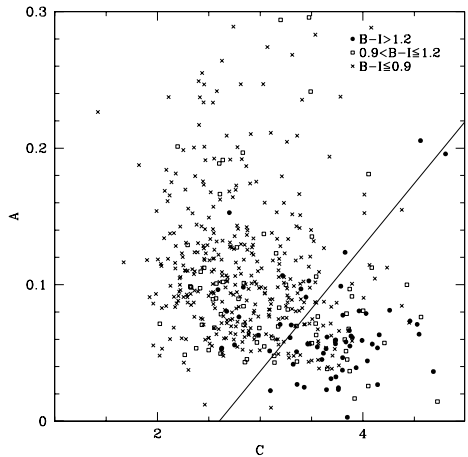,width=6.5cm} 
\vspace*{0.25cm}  
\caption{Asymmetry-concentration diagram of the spectroscopic redshift sample, as a function of the rest-frame color index $(B-I)_{RF}$.}
\label{fig: AC color}
\end{figure}

We studied the dependence on the color of the $(A, C)$ diagram.
 Absolute magnitudes and color indices were determined by fitting spectral energy distribution
(SED) templates at fixed spectroscopic redshift on COMBO-17 multi-color
data (\cite{wolf04}) using \textit{Le
Phare}\footnote{www.lam.oamp.fr/arnouts/LE\_PHARE.html} (S. Arnouts \&
O. Ilbert).
RF color index and location in the $(A, C)$ diagram seem
correlated (Figure \ref{fig: AC color}). The reddest objects are
mostly the bulge-dominated galaxies; and the bluest are mostly the disk-dominated galaxies. 
It clearly appears  some unexpected population of red (or blue)
galaxies in the disk-dominated area (or respectively bulge-dominated
area).  The red galaxies in the disk-dominated area could be simply explained by 
extinction. The nature of the blue population in the
bulge-dominated area remains unclear (see later). Both informations, morphology and color,
are complementary to discriminate between two different populations in
the bulge-dominated area. It is crucial to interpret our results in
the galaxy evolution. 

\section{Evolution with the redshift}

A large number of galaxies is required to perform a study of the
galaxy evolution. In order to improve our statistics, we used
hereafter the huge photometric redshift sample.

We selected the 817 galaxies that are brighter than
$M_{AB}(B)=-20$ in the range of redshift [0.15-1.1], in order to avoid
selection effects in apparent magnitude ($I_{AB}<24$). In Figure \ref{fig:
ratio MB<-20}, on the left,  we plotted the proportion of bulge- and
disk-dominated galaxies as a function of the redshift. We did not see any evident trend in the variation
with $z$ of the proportion of morphological types. This result suggests
that the majority of early-type galaxies was in place at $z\sim1$, in
agreement with the works of \cite{schade99}, \cite{kajisawa} and
\cite{conselice05}. The only general trend is for the bluest bulge-dominated galaxies ($B-I<0.9$). These galaxies are more frequent (or brighter)
at high redshift: their rate monotonically increases from $z\sim0.4$
toward $z\sim1$.
The proportion of each
type is sensitive to the presence of large
structures in the field. At $z\sim0.7$ which is a high-density region,
the rate of red bulge-dominated galaxies is particularly high, reaching
38\% of the galaxies at this redshift. Our results are consistent with
morphology-density relations (e.g. \cite{dressler97}). The
high rate of bulge-dominated galaxies in dense environment is
expected in the hierarchical picture since there is an higher
probability of merging in the large structure of black matter.
 However a volume limited sample has the inconvenient to
throw out the faintest galaxies of the sample, which is an
inconvenient to study the evolution of this faint blue bulge-dominated
population.

\begin{figure}[htbp]
\vspace*{1.25cm}  
\centering
\begin{tabular}{cc}
\epsfig{figure=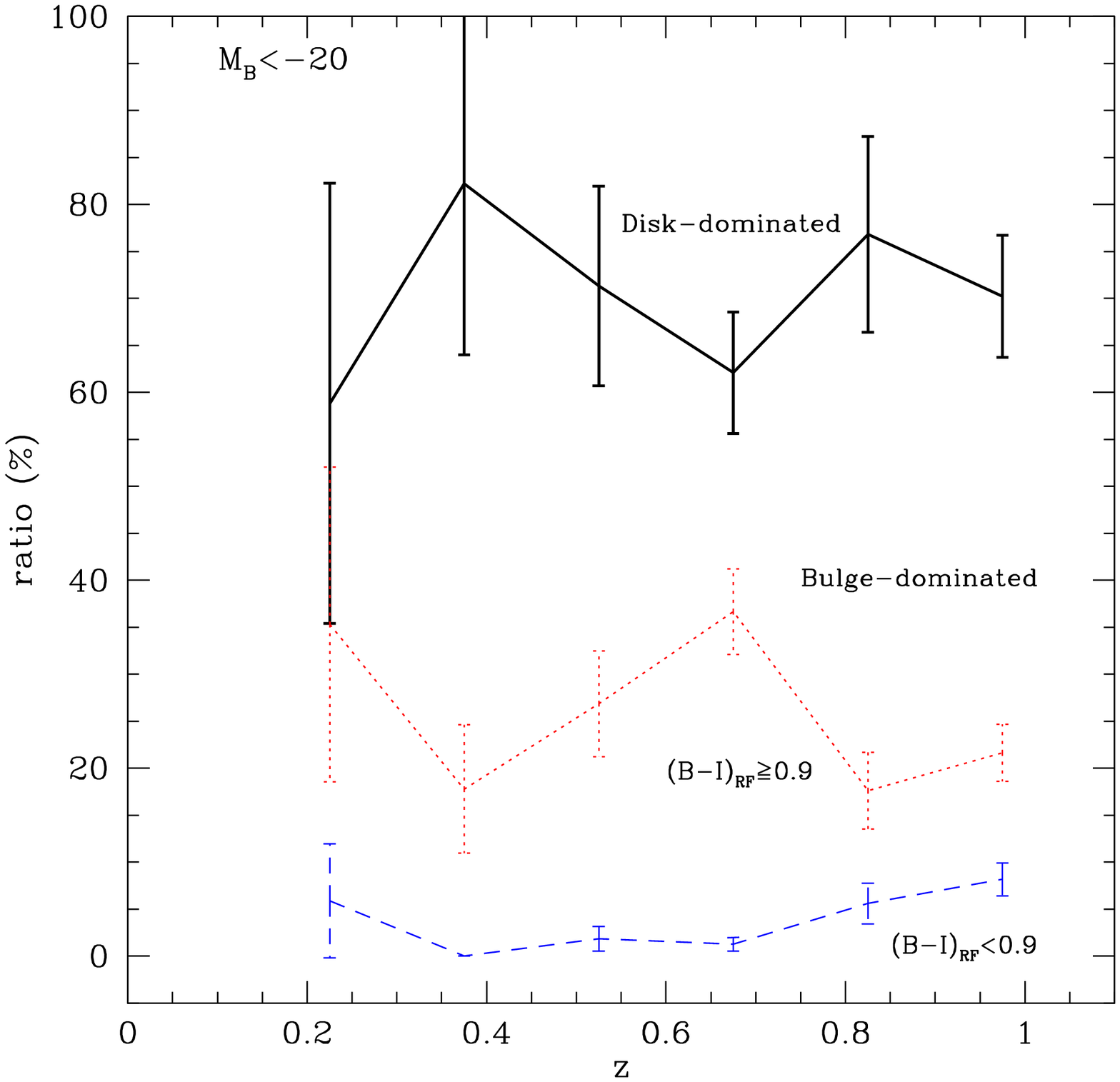,width=7cm} & \epsfig{figure=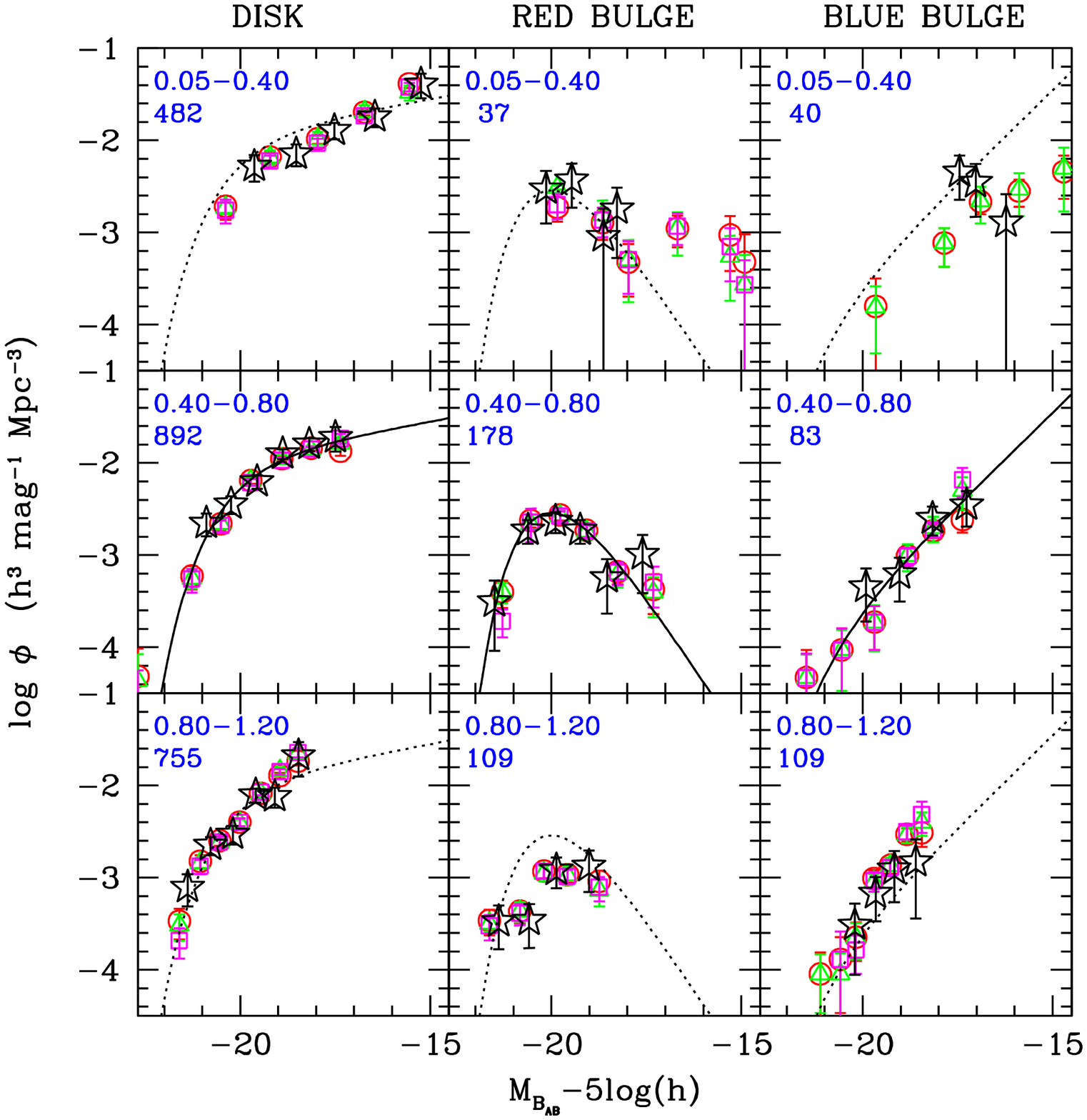,width=7cm} \\ 
\end{tabular}
\vspace*{0.25cm}  
\caption{Evolution of the proportion of bulge- and disk-dominated
galaxies brighter than $M_{AB}(B)=-20$, at $0.15<z<1.1$. 
The bulge-dominated population is divided into a blue (RF $B-I<0.9$) population (dashed line) and a red (RF $B-I\ge0.9$) population (dotted line). Each point is
computed on an interval of redshift ${\Delta}z=0.15$. The error bars
at 1$\sigma$ are poissonnian.}
\label{fig: ratio MB<-20}
\end{figure}

We derived the LF per morphological type using the Algorithm for Luminosity Function (ALF)
described in \cite{ilbert}.
The Figure \ref{fig: ratio MB<-20} shows on the right the evolution of the disk-, red bulge- and
blue bulge-dominated populations from $z=0.05$ up to $z=1.2$.  The results
obtained with the photometric redshift sample are fully in agreement
with the results obtained with the spectroscopic redshift sample (see
the open stars of Figure \ref{fig: ratio MB<-20}, on the right). This comparison insured us that
our measurement of the LF evolution is not due to systematic trends in
the photometric redshift estimate.

The LF of the disk-dominated galaxies remains consistent over all the
redshift range. The slope of the disk-dominated galaxies is also in
agreement with local values obtained by \cite{Marinoni99}
 or by
\cite{Nakamura03}. We found a very small
evolution of the disk-dominated population although irregular galaxies which are included in the disk-dominated population are expected to evolve strongly with $z$.
The density of the red bulge-dominated population decreases at high
redshift $z=0.8-1.2$.  The evolution of the LF for the red bulge-dominated
population is opposite to the increase in density observed by
\cite{Cross04} between $z=0.5-0.75$ and $z=0.75-1$ but is in
agreement with the result obtained by \cite{Ferreras05} on the
same field. However, the observed evolution of the red bulge-dominated
population remains small in comparison to the strong decrease in
density of the elliptical galaxies measured by \cite{Wolf03}
using a photometric type. Our results show rather a population of
Ell/S0 galaxies already in place at $z\sim 1$.

The LF slope of the blue bulge-dominated population remains extremely
steep in all the redshift bins. This population strongly evolves. To
quantify the LF evolution, we first set the $M^*-\alpha$ parameters to
the $z=0.4-0.8$ values and look for an evolution in $\Phi^*$. We
measure an increase in density of a factor 2.4 between $z=0.4-0.8$
and $z=0.8-1.2$. Using the same procedure, we set the $\alpha -
\Phi^*$ parameters to the $z=0.4-0.8$ values and look for an evolution
in $M^*$. We measure a brightening of 0.7 magnitude between
$z=0.4-0.8$ and $z=0.8-1.2$. The same trend is also present between
$z=0.05-0.4$ and $z=0.4-0.8$. \\
The nature of these blue, faint bulge-dominated galaxies must be investigated. It is unlikely than a large fraction of
this blue bulge-dominated population are misclassified spiral at high
redshift since the visual inspection of the UV rest-frame image at
$z\sim 1$ should have clearly shown star formation region in the
spiral arms. If a large fraction of these blue bulge-dominated
galaxies are misclassified spirals at $z<0.7$ as claimed by
\cite{Ferreras05}, the observed evolution of the blue
bulge-dominated galaxies is even stronger.\\

\section{Conclusions} 
\begin{enumerate}
\item Using multi-band imaging of HST/ACS, we were able to build a reliable morphological classification of galaxies in the CDFS. An asymmetry-concentration diagram in rest-frame B band allowed us to distinguish bulge- and disk-dominated galaxies. The study of the correlation between morphology and color highlights that both informations are complementary: even if the link between morphology and color is very strong, we found unexpected blue bulge-dominated galaxies whose nature is still unclear.
\item Using a $M_{AB}(B)<-20$ selection within the interval of redshifts [0.15-1.1], we studied the evolution of the proportion of morphological types with the redshift. We did not find any significant general trend in the variation with $z$ of the proportion of disk- and bulge-dominated galaxies, in agreement with works of \cite{schade99}, \cite{kajisawa}, \cite{conselice05} who suggested that the majority of early-type galaxies was already in place at $z\sim1$.
\item The variation of the proportion of morphological types appeared to be very sensitive to the presence of large structures. At $z\sim0.7$ where a wall of galaxies is detected in the redshift distribution, the proportion of bright red bulge-dominated galaxies reached a peak of 38\%, to about 20\% at $z=0.4$ and at $z\sim0.8$. This results is consistent with the hierarchical scenario of formation of early-type galaxies.
\item We derived the rest-frame $B$-band LF per morphological type up to
$z=1.2$. We obtained a strong dependency of the LF shape on the
morphological type. The LF of the disk-dominated population has a steep slope, which does not change significantly with the redshift. The LF of the
bulge-dominated population is the combination of two populations: a
red, bright population (68\% of the bulge-dominated sample) with a strong decreasing slope; and a blue
population of more compact galaxies (32\% of the bulge-dominated
sample) dominating the fain-end of the LF. 
\item We measured an increase in density of red bulge-dominated population
with the age of the Universe. The observed evolution of the red
bulge-dominated LF could be related to the building-up of massive
elliptical galaxies in a hierarchical scenario, formed by merging and
accretion of smaller galaxies. We observe a very strong evolution of the blue
bulge-dominated population corresponding to a brightening of 0.7
magnitude (or an increasing in density by a factor 2.4) between $z
\sim 0.6$ and $z \sim 1$.
\end{enumerate}
Our analysis clearly shows the next steps toward a strong constraint on
the scenarios of galaxy formation: an increase of the covered area to
limit the cosmic variance which will be possible with the COSMOS
survey, the development of quantitative methods to discriminate more
morphological classes in particular merger and irregular galaxies, the
necessity to combine morphological and spectral classifications.


\vfill 
\end{document}